\documentclass[12pt,letterpaper]{article}
\RequirePackage{natbib}
\usepackage{anysize}
\usepackage{amsmath}
\usepackage{amssymb}
\RequirePackage{amsfonts}
\usepackage{times}
\usepackage{enumerate}
\usepackage{multirow} 
\usepackage{url}
\usepackage{graphicx} 
\usepackage{version}
\usepackage{setspace}
\usepackage{mathbbol}

\newcommand{\keywords}{\text{KEY WORDS:}}

\includeversion{unblinded}\excludeversion{blinded}

\title{Monte Carlo convergence of rival samplers}

\begin{unblinded}
\author{Nicholas A. \textsc{Heard} and Melissa J. M. \textsc{Turcotte}
}
\end{unblinded}

\newcommand{\argmax}{\arg \! \max}
\newcommand{\ns}{n}
\newcommand{\jsd}{JSD}
\newcommand{\E}{\mathbb{E}}
\newcommand{\KL}{\mathrm{KL}}

\begin{document}
\maketitle 

\begin{abstract}
It is often necessary to make sampling-based statistical inference
about many probability distributions in parallel. Given a finite
computational resource, this article addresses how to optimally divide
sampling effort between the samplers of the different
distributions. Formally approaching this decision problem requires
both the specification of an error criterion to assess how well each
group of samples represent their underlying distribution, and a loss
function to combine the errors into an overall performance score. For
the first part, a new Monte Carlo divergence error criterion based on
Jensen-Shannon divergence is proposed. Using results from information
theory, approximations are derived for estimating this criterion
for each target based on a single run, enabling adaptive sample size
choices to be made during sampling.



\end{abstract}


\section{Introduction}

Let $X_{1},X_{2},\ldots$ be a sequence of random samples obtained from
an unknown probability distribution $\pi$. The corresponding random
measure from $n$ samples is the Monte Carlo estimator of $\pi$,
\begin{equation}
\hat{\Pi}_\ns(B)=\frac{1}{\ns}\sum_{i=1}^{\ns} \delta_{X_i}(B), ~~B\subseteq \mathcal{X}, \label{eq:empirical_measure_}
\end{equation}
where $\mathcal{X}$ is the support of $\pi$. The random measure
\eqref{eq:empirical_measure_} is a maximum likelihood estimator of
$\pi$ and is consistent: for all $\pi$-measurable sets $B$,
$
\lim_{\ns\to\infty} \hat{\Pi}_\ns(B) = \pi(B).
$

Sometimes estimating the entire distribution $\pi$ is of intrinsic
inferential interest. In other cases, this may be desirable if there
are no limits on the functionals of $\pi$ which might be of future
interest. Alternatively, the random sampling might be an intermediary
update of a sequential Monte Carlo sampler 
no
for which it is desirable that the samples represent the current
target distribution well at each step \citep{delmoral06}.


Pointwise Monte Carlo errors are inadequate for capturing the overall
rate of convergence of the realised empirical measure $\hat{\pi}_\ns$
to $\pi$. This consideration is particularly relevant if $\pi$ is an
infinite mixture of distributions of unbounded dimension: In this case
it becomes necessary to specify a degenerate, fixed dimension function
of interest before Monte Carlo error can be assessed
\citep{sisson_fan07}. This necessity is potentially undesirable,
since the assessment of convergence will vary depending on which
function is selected and that choice might be somewhat arbitrary.

The work presented here considers sampling multiple target
distributions in parallel. This scenario is frequently encountered in
real-time data processing, where streams of data pertaining to
different statistical processes are collected and analysed in fixed
time-window updates. Decisions on how much sampling effort to allocate
to each target will be made sequentially, based on the apparent
relative complexity of the targets, as higher-dimensional, more
complex targets intuitively need more samples to be well
represented. The complexities of the targets will not be known
\textit{a priori}, but can be estimated from the samples which have
been obtained so far. As a consequence, the size of the sample $\ns$
drawn from any particular target distribution will be a realisation of
a random variable, $N$, determined during sampling by a random
stopping rule governed by the history of samples drawn from that
target and those obtained from the other targets.

To extend the applicability of Monte Carlo error to entire probability
measures, the following question is considered: If a new sample of
random size $N$ were drawn from $\pi$, how different to
$\hat{\pi}_\ns$ might the new empirical measure be? If repeatedly
drawing samples in this way led to relatively similar empirical
measures, this suggests that the target is relatively well represented
by $N$ samples; whereas if the resulting empirical measures were very
different, then there would be a stronger desire to obtain a
(stochastically) larger number of samples. To formally address this
question, a new \textit{Monte Carlo divergence error} is proposed to
measure the expected distance between an empirical measure and its
target.

Correctly balancing sample sizes is a non-trivial
problem. Apparently sensible, but \textit{ad hoc}, allocation
strategies can lead to extremely poor performance, much worse than
simply assigning the same number of samples to each target. Here, a
sample-based estimate of the proposed Monte Carlo divergence error of
an empirical measure is derived; these errors are combined across
samplers through a loss function, leading to a fully-principled,
sequential sample allocation strategy. 

Section \ref{sec:mc_convergence} formally defines and justifies Monte
Carlo divergence as an error criterion. Section
\ref{sec:rival_samplers} examines 
two different loss functions for combining sampler errors into a
single performance score. Section \ref{sec:other_methods} introduces
some alternative sample size selection strategies; some are derived by
adapting related ideas in the existing literature, and some are
\textit{ad hoc}. The collection of strategies are compared on
univariate and variable dimension target distributions in Section
\ref{sec:examples} before a brief discussion in Section
\ref{sec:discussion}.

\section{Monitoring Monte Carlo convergence}\label{sec:mc_convergence}


In this section the rate of convergence of the empirical distribution to the target 
will be assessed by information theoretic criteria. In information
theory, it is common practice to discretise distributions of any
continuous random variables \citep[see][]{paninski03}. Without this
discretisation (or some alternative smoothing) the intersection
of any two separately generated sets of samples would be empty, and
distribution-free comparisons of their empirical measures would be
rendered meaningless:
For example, the Kullback-Leibler divergence between two independent
realisations of \eqref{eq:empirical_measure_} will be always be
infinite.

When a target distribution relates to that of a continuous random
variable, a common discretisation of both the empirical measure and
notionally the target will be performed. For the rest of the article,
both $\hat{\pi}$ and $\pi$ should be regarded as suitably discretised
approximations to the true distributions when the underlying variables
are continuous. When there are multiple distributions, the same
discretisation will be used for all distributions.
For univariate problems a large but finite grid with fixed spacing
will be used to partition $\mathcal{X}$ into bins; for mixture
problems with unbounded dimension, the same strategy will be used for
each component of each dimension, implying an infinite number of bins.
Later in Section \ref{sec:bin_width}, consideration will be given to
how the number of bins for each dimension should be chosen.

\subsection{Monte Carlo divergence error}\label{sec:mc_divergence}

For a discrete target probability distribution $\pi$, let
$\hat{\Pi}_n$ be the estimator \eqref{eq:empirical_measure_} for a
prospective sample of size $\ns$ to be drawn from $\pi$, and let
$\hat{\Pi}$ be the same estimator when the sample size is a
random stopping time.

For $\ns\geq 1$, the Monte Carlo divergence error of the estimator
$\hat{\Pi}_n$ will be defined as
\begin{equation}
e_{\KL,\ns}=H(\pi) - \E\{H(\hat{\Pi}_\ns)\}, \label{eq:e_KL_n}
\end{equation}
where $H$ is Shannon's entropy function; recall that if
$p=(p_1,\ldots,p_K)$ is a probability mass function,
$
H(p)=-\sum_{i=1}^K p_i \log(p_i).
$
Note that $H(\hat{\Pi}_\ns)$ is the maximum likelihood estimator of
$H(\pi)$. The Monte Carlo divergence error
$e_{\KL,\ns}$ has a direct interpretation: it is \textit{the expected
  Kullback-Leibler divergence of the empirical distribution of a
  sample of size $\ns$ from the target $\pi$}, and therefore provides
a natural measure of the adequacy of $\hat{\Pi}_n$ for estimating
$\pi$.
 
The Monte Carlo divergence error of the estimator $\hat{\Pi}$ when
$\ns$ is a random stopping time is defined as the expectation of
$e_{\KL,\ns}$ with respect to the stopping rule, or equivalently
\begin{equation}
e_{\KL}=H(\pi) - \E\{H(\hat{\Pi})\}, \label{eq:e_KL}
\end{equation}
where the expectation in \eqref{eq:e_KL} is now with respect to both
$\pi$ and the stopping rule. This more general definition of Monte
Carlo divergence error should be interpreted as \textit{the expected
  Kullback-Leibler divergence of the empirical distribution of a
  sample of random size from the target $\pi$}.

To provide a sampling based justification for this definition of Monte
Carlo divergence error, for $M>1$ consider the empirical distribution estimates
$\hat{\pi}^{1},\dots,\hat{\pi}^{M}$ which would be obtained from
$M$ independent repetitions of sampling from $\pi$, where the
sample size of each run is a random stopping time from the same rule.

The Jensen-Shannon divergence \citep{lin_wong90,lin91} of $\hat{\pi}^{1},\dots,\hat{\pi}^{M}$,
\begin{equation}
\jsd(\hat{\pi}^{1},\dots,\hat{\pi}^{M})=
H\left(\frac{1}{M}\sum_{j=1}^M  \hat{\pi}^{j}\right) -\frac{1}{M}\sum_{j=1}^M  H(\hat{\pi}^{j}),
\label{eq:JSD}
\end{equation}
measures the variability in these distribution estimates by
calculating their average Kullback-Leibler divergence from the closest
dominating measure, which is their average. The Jensen-Shannon
divergence is a popular quantification of the difference between
distributions, and its square root has the properties of a metric on
distributions \citep{endres03}.

Just as Monte Carlo variance is the limit of the sample
variance of $M$ sample means as $M\to\infty$, the Monte Carlo divergence error defined
in \eqref{eq:e_KL} is easily seen to be the limit of \eqref{eq:JSD} as
$M\to\infty$: By the strong law of large numbers,
$
\lim_{M\to\infty} \frac{1}{M}\sum_{j=1}^M  \hat{\pi}^{j} = \pi,
$
and 
$
\lim_{M\to\infty} \frac{1}{M} \sum_{j=1}^M  H(\hat{\pi}^{j}) = \E[H(\hat{\Pi})],
$
the expected entropy of a Monte Carlo distribution estimate from one
of the runs.  It follows that \eqref{eq:JSD} is a biased but
consistent estimate of \eqref{eq:e_KL}.

Finally, it should be noted that there is a second interpretation of
the Monte Carlo divergence error: $e_{\KL}$ is also the \emph{negative
  bias of the maximum likelihood estimator of the entropy of $\pi$}
given a random sample. In the next section it will be shown that this
alternative interpretation is very useful, since it leads to a
mechanism for estimating $e_{\KL}$ from a single sample.

\subsection{Estimating Monte Carlo divergence error}\label{sec:mc_divergence_estimation}

While $H(\hat{\Pi}_\ns)$ is the maximum likelihood estimator of
$H(\pi)$, it is known to be a negatively biased since $H$ is a concave
function \citep{Miller_1955}. Various approximate bias corrections for
$H(\hat{\Pi}_\ns)$ have been proposed in the information theory
literature, and these correction terms can serve here as approximately
unbiased estimates of $e_{\KL,n}$. Furthermore, note that any unbiased
estimate of $e_{\KL,\ns}$ is also an unbiased estimate of $e_{\KL}$,
the error under the random stopping rule.

Given a sample of size $\ns$, the popular Miller-Madow method
estimates the negative bias of the maximum likelihood estimate
$H(\hat{\pi})$ to be $(K-1)/(2\ns)$, where $K$ is the number of
nonempty bins in $\hat{\pi}$. This estimate proves to be too crude for
the purpose here of estimating $e_{\KL,n}$, since any two
distributions with the same number of represented bins would be
estimated to have the same divergence, regardless of how uniform the
corresponding bin probabilities might be.

An improvement on the Miller-Madow estimate was provided by
\cite{Grassberger1988,Grassberger2003},
\begin{equation}
\hat{e}_{\KL, \ns}=\frac{1}{\ns}\sum_{i=1}^K  \phi(n_i),
\label{eq:ehat_KL}
\end{equation}
where $n_i$ is the number of samples in the $i$th nonempty bin, such
that $\sum_{i=1}^K n_i=\ns$, and 
$
\phi(n_i)= n_i\{\log(n_i)-\psi(n_i)\},
$
where $\psi$ is the digamma function.
In this work, \eqref{eq:ehat_KL} will provide an approximately
unbiased estimate of $e_{\KL,n}$, the expected Kullback-Leibler
divergence of the empirical distribution from the target, based on a
single run of the sampler.

\subsubsection{Efficient calculation}\label{sec:efficient_calculation}
Calculation of \eqref{eq:ehat_KL} during sampling can be updated at
each iteration very quickly, using the following equations. Let $i'$
be the bin in which the $n$th observation falls. Then
\begin{equation}
\hat{e}_{\KL, \ns}=\frac{(n-1)\hat{e}_{\KL,
    \ns-1}+\Delta^{1}\phi(n_{i'}-1)}{\ns},
\label{eq:ehat_KL_update}
\end{equation}
where the forward difference operator 
$\Delta^1
\phi(n_{i'}-1)=\phi(n_{i'})-\phi(n_{i'}-1)$.

Besides estimating the current Monte Carlo divergence error of a
distribution estimate after $\ns$ samples from $\pi$, it will also be of
interest to estimate the expected reduction in error that would be
achieved from obtaining one more sample,
\begin{equation}
\delta_{\KL,n} = e_{\KL,n+1}-e_{\KL,n}. \label{eq:de_KL}
\end{equation}

To estimate this quantity it is now necessary to assume that samples
are drawn approximately independently (perhaps via thinned MCMC), and
that the probability of the new sample falling into the $i$th bin is
approximated by the empirical, maximum likelihood estimate $n_i/\ns$.
%
Then the expected reduction in error from a further sample can be estimated
as
\begin{eqnarray}
\hat{\delta}_{\KL, \ns}&=&\frac{1}{\ns} \sum_{i=1}^K \sum_{j=0}^{1} n_i^j(\ns-n_i)^{1-j}\left\{\frac{\phi(n_i)}{\ns} - \frac{\phi(n_i+j)}{\ns+1} \right\}\nonumber\\
&=&\frac{1}{\ns(\ns+1)} \sum_{i=1}^K  (n_i+1)\phi(n_i) - n_i\phi(n_i+1).
\label{eq:dehat_KL}
\end{eqnarray}
Again considering this calculation iteratively, if the $n$th
observation falls into bin $i'$ then
\begin{eqnarray}
\hat{\delta}_{\KL, \ns}=\frac{(\ns-1)\ns\hat{\delta}_{\KL, \ns-1} - n_{i'}\Delta^2\phi(n_{i'}-1)}{\ns(\ns+1)},
\label{eq:dehat_KL_update}
\end{eqnarray}
where the second forward difference $\Delta^2
\phi(n_{i'}-1)=\phi(n_{i'}+1)-2\phi(n_{i'})+\phi(n_{i'}-1)$.

\subsubsection{Alternative formulations of bias estimation}

It should be noted that further refinements (additive terms) to the
bias estimate of \eqref{eq:ehat_KL} are provided by
\cite{Grassberger2003}, such as
\begin{equation}
\phi(n_i)= n_i\{\log(n_i)-\psi(n_i)\} - \frac{{-1}^{n_i}}{n_i+1}.
\end{equation}
However, these additional terms, which arise as part of a second order
approximation of an integral, are unstable, oscillating between
positive and negative values. In this context, without careful
treatment such terms can incorrectly suggest that the expected error
might very slightly increase by taking further samples, which in
practice is not true but would make obtaining further samples seem
undesirable. Furthermore, due to their oscillating sign, these terms
do not affect the overall drift of the function, which will be the
quantity of longer term interest when deciding whether more sampling
effort should be afforded.

\section{Rival samplers}\label{sec:rival_samplers}

Consider $m$ probability distributions $\pi_1,\ldots,\pi_m$.  Suppose
random samples are to be drawn from a sampler for each $\pi_j$, and
that the empirical distributions of the samples will eventually serve as
approximations of the corresponding target distributions.

Given a fixed computational resource, which might simply correspond to
a final total number of random samples $\ns$, the decision problem to
be addressed is how best to divide those $\ns$ samples between the $m$
samplers. That is, how sample sizes $\ns^{(1)},\ldots,\ns^{(m)}$ for
the $m$ targets should be chosen subject to the constraint
$\sum_{j=1}^m \ns^{(j)} = \ns$. The samplers can be viewed as \textit{rivals}
to one another for the same fixed computational resource.

Without this or a similar constraint, the problem would be ill-posed:
for all $j$, $\ns^{(j)}$ should be chosen to be as large as possible,
since Monte Carlo errors are monotonically decreasing with sample
size. A constraint is required for any sample size choice to be
practically meaningful. In contrast, results which establish a
minimum sample size for which errors should fall within a (typically
arbitrary) desired level of precision are theoretically interesting,
but are perhaps best viewed in the reverse direction in this context;
given the inevitable usage of the maximum allowable computation time,
understanding the level of error this limit implies.

The default choice is for equal samples sizes, $\ns^{(j)} = \ns/m$,
but such an approach disregards any differences in the
\textit{complexities} of the target distributions, which in general
could be arbitrarily different. The aim of this work is to adaptively
determine how much sampling effort should be afforded to each
sampler. The preceding section established a general method for
assessing the error of each sampler. The choice of how to balance
sample sizes between the samplers will be made according to a loss
function for combining those errors.

\subsection{Loss functions for Monte Carlo errors}\label{sec:loss}
Suppose that the decision to assign $n^{(j)}$ of the total $\ns$
samples to the sampler of $\pi_j$ implies a Monte Carlo error level
$e^{(j)}_{n^{(j)}}$ for that target. The specific definition of this
Monte Carlo error can be left open; for example, this might be the
usual Monte Carlo error of a point estimate; or if interest lies in
summarising the whole distribution, the Monte Carlo divergence error
criterion \eqref{eq:e_KL}.

Two natural alternative loss functions for combining the individual
errors $e^{(j)}_{n^{(j)}}$ into an overall performance error are
considered here. One possibility is that utility could be
derived from controlling the maximum error of
the $m$ samplers,
suggesting an (expected) loss function
\begin{equation}
\mathcal{L_{\mathrm{max}}}(\ns^{(1)},\ldots,\ns^{(m)})=\max_{j\in\{1,\ldots,m\}}e^{(j)}_{n^{(j)}}.
\label{eq:max_loss}
\end{equation}
This form of loss function could be applicable in financial trading,
for example, where exposure to the worst loss could be
unlimited. Alternatively it might be important to control the average
error across the samplers, suggesting a different loss function of the
form
\begin{equation}
\mathcal{L_{\mathrm{ave}}}(\ns^{(1)},\ldots,\ns^{(m)})=\frac{1}{m}{\sum_{j=1}^m e^{(j)}_{n^{(j)}}}.
\label{eq:ave_loss}
\end{equation}
This form could be applicable in portfolio trading, where exposure to
loss is spread across the composite stocks. To illustrate the
difference between these two loss functions, consider estimating the
means of two distributions with known variances $\sigma^2_1$,
$\sigma^2_2$, with error measured by the Monte Carlo variance of the
sample means; the optimal ratio of sample sizes, $n^{(1)}/n^{(2)}$,
would be given by the ratio of the variances $\sigma^2_1/\sigma^2_2$
under $\mathcal{L_{\mathrm{max}}}$, and by the ratio of the standard
deviations $\sigma_1/\sigma_2$ under
$\mathcal{L_{\mathrm{ave}}}$. Therefore care should be exercised in
specifying the required form of loss function. Other choices, or
indeed linear combinations of these two losses, could be examined.

\subsection{Sequential allocation of samples}\label{sec:sequential}
Away from the stylised example of the previous section, in general it
is more likely that little will be known \textit{a priori} about the
target distributions being sampled. Instead, the aim will be to
dynamically decide, during sampling, which samplers should be afforded
more sampling effort, conditional on the information learned so far
about the targets. A sequential decision approach is taken. Having taken
$n'<\ns$ samples, with $\ns^{(j)}$ of these allocated to the $j$th
sampler, the decision problem is to choose from which sampler to draw
the $(n'+1)$th sample, such that the chosen loss function of the
estimated Monte Carlo errors of the samplers
$\{\hat{e}^{(j)}_{n^{(j)}}\}$ is minimised.

In this sequential setting, the operational difference between the
loss functions $\mathcal{L_{\mathrm{max}}}$ and
$\mathcal{L_{\mathrm{ave}}}$ becomes clearer. If the aim is to
minimise $\mathcal{L_{\mathrm{max}}}$, then the optimal decision for
allocating one more sample is to allocate it to the sampler with the
highest estimated error,
\begin{equation}
\argmax_j ~ \hat{e}^{(j)}_{n^{(j)}},\label{eq:max_rule}
\end{equation}
since error is a decreasing function of sample size. Alternatively, if
minimising $\mathcal{L_{\mathrm{ave}}}$ then the new sample should be
allocated to the sampler for which the estimated \textit{decrease} in
error is highest,
\begin{equation}
\argmax_j ~ \hat{\delta}^{(j)}_{n^{(j)}},\label{eq:ave_rule}
\end{equation}
since this will minimise the overall expected sum.

These sequential decision rules are myopic, looking only one step
ahead. There are three reasons why this is preferred; first,
considering optimal sequences of future allocations leads to a
combinatorial explosion unsuitable for a method intended for
optimising the use of a fixed computational resource; second, the
final number of samples may even be unknown; third, the estimated
error or expected change in error under the Monte Carlo divergence
criterion \eqref{eq:ehat_KL}, \eqref{eq:dehat_KL} can be updated very
efficiently via \eqref{eq:ehat_KL_update} and
\eqref{eq:dehat_KL_update}: after one more sample, only one bin count
$n_i^{(j)}$ for one sampler $j$ will have changed.

Any sequential sampling allocation scheme which depends on the
outcomes of the random draws will imply a random stopping rule for the
number of samples eventually allocated to each sampler. This adds an
extra complication, since some stopping rules will introduce bias into
Monte Carlo estimates \citep{mendo_hernando06}. Here this bias arises
if the first samples taken from a target distribution have a
particularly low estimated Monte Carlo error, as this will cause the
other rival samplers to share all of the remaining samples; without
corrective action, this phenomenon causes Monte Carlo estimators to be
biased towards estimates of this character.

When the Monte Carlo error is the divergence measure \eqref{eq:e_KL},
low error estimates correspond to low entropy empirical measures,
which can spuriously arise if the first random samples happen to fall
into the same bin. Therefore, to eradicate this bias, a minimum number
of samples $\ell^{(j)}$ is recommended for each target distribution,
to prevent degenerate sample sizes. To examine stability, these minima
can be chosen in increasing steps until the resulting samples sizes
converge. For the examples in Section \ref{sec:examples}, due to the
relatively fine grid used for binning samples 
it was enough to set $\ell^{(j)}=500$ to obtain convergence.

\subsection{Algorithm: Rival sampling}\label{sec:alg}
The full algorithm for sequential sampling from rival target
distributions to minimise estimated loss is now presented. For $m$
samplers of target distributions $\pi_1,\ldots,\pi_m$, let
$\ell^{(j)}\geq 1$ be the minimum number of samples that should be
drawn from $\pi_j$. Let
$\mathcal{L}\in\{\mathcal{L_{\mathrm{max}}},\mathcal{L_{\mathrm{ave}}}\}$
be the chosen loss function for combining Monte Carlo errors across
the samplers.

 The algorithm proceeds as follows:
\begin{enumerate}
\item Initialise--- For $j=1,\ldots,m$:
\begin{enumerate}
\item Draw $\ell^{(j)}$ samples from $\pi_j$ and calculate
  $\hat{\pi}_j$, the binned empirical estimate of $\pi_j$ assuming $K$
  bins in each dimension; let $K_j$ be the number of non-empty bins in $\hat{\pi}_j$,
  and $\ns^{(j)}_1,\ldots,\ns^{(j)}_{K_j}$ be the corresponding bin
  counts; set $\ns^{(j)}=\ell^{(j)}$.

\item  \noindent If $\mathcal{L} = \mathcal{L_{\mathrm{max}}}$: calculate the
  divergence estimate for the $j$th sampler, $\hat{e}_{
    \ns^{(j)}}^{(j)}$, using \eqref{eq:ehat_KL};

  \noindent else if $\mathcal{L} = \mathcal{L_{\mathrm{ave}}}$:
  calculate the estimated increment in divergence for the $j$th
  sampler, $\hat{\delta}^{(j)}_{\ns^{(j)}}$, using \eqref{eq:dehat_KL}.
\end{enumerate}

\item Iterate--- Until the available computational resource is exhausted:
\begin{enumerate}
\item \noindent If $\mathcal{L} = \mathcal{L_{\mathrm{max}}}$: set $j^\ast =
\argmax_j ~ \hat{e}^{(j)}_{n^{(j)}}$;

\noindent else if  $\mathcal{L} = \mathcal{L_{\mathrm{ave}}}$: set $j^\ast =
\argmax_j ~ \hat{\delta}^{(j)}_{n^{(j)}}$.

\item Sample one new observation from $\pi_{j^\ast}$. Set
  $\ns^{(j^\ast)}=\ns^{(j^\ast)}+1$. Let $i$ be the bin into which the
  new observation falls. 
  Set $\ns^{(j^\ast)}_i=\ns^{(j^\ast)}_i+1$. If bin $i$ was previously
  empty, set $K_{j^\ast}=K_{j^\ast}+1$.

\item Update $\hat{e}^{(j)}_{n^{(j)}}$ or
  $\hat{\delta}^{(j)}_{n^{(j)}}$ using \eqref{eq:ehat_KL_update} or
  \eqref{eq:dehat_KL_update} respectively.

\end{enumerate}
\end{enumerate}

\subsection{Choosing a bin width for discretisation}\label{sec:bin_width}
The algorithm of Section \ref{sec:alg} requires a method of
discretising samples from continuous distributions. (For simplicity, a
fixed bin width can be assumed for each dimension of a multivariate
distribution.)
The following observations offer some insight for what makes a good
bin width in this context. In the limit of the bin width going to
zero, the binned empirical distribution after $\ns$ independent draws
would have $\ns$ non-empty bins each containing one
observation. Although the identity of those bins would vary across
samplers, these empirical distributions would be indistinguishable in
terms of both entropy and \eqref{eq:e_KL}; so each sampler would be
allocated the same sample size. In the opposite limit of the bin width
becoming arbitrarily large, all samples of the same dimension would
fall into the same bin. For fixed dimension problems, this would mean
all sample sizes would be equal, and otherwise in transdimensional
problems the strategies would simplify to working with marginal
distributions of the dimension, which reduces the potential diversity
of sample sizes. So a good bin width would lie well within these two
extremes, ideally maximising the resulting differences in sample
sizes. That is, a good bin width should distinguish well the varying
complexity of the different targets. Further to these observations,
the next section suggests a novel maximum likelihood approach for
determining an optimal number of bins, which 
could be deployed adaptively or using the initial $\ell^{(j)}$ samples
drawn from each target.

\subsubsection{Maximum likelihood bin width estimation for Bayesian histograms}\label{sec:bayesian_bin_width}
Consider a regular histogram of $K$ equal width bins on the interval
$[a,b]$, and let $p=(p_1,\ldots,p_K)$ be
the bin probabilities. The Bayesian formulation of this histogram
\citep{LEONARD01081973} treats the probabilities $p$ as unknown, and a
conjugate Dirichlet prior distribution based on a Lebesgue base
measure with confidence level $\alpha$ suggests
$p\sim\mathrm{Dirichlet}(\alpha(b-a)/K\cdot\boldsymbol{1}^\mathrm{T})$. For
$n$ samples, the marginal likelihood of observing bin counts
$n_1,\ldots,n_K$ under this model is
\begin{equation}
\Gamma\{\alpha(b-a)\}/[\Gamma\{\alpha(b-a)+n\}\Gamma\{\alpha(b-a)/K\}^K\{(b-a)/K\}^n]\prod_{i=1}^K\Gamma\{\alpha(b-a)/K+n_i\}.\label{eq:density_ml}
\end{equation}
Using standard optimisation techniques, identifying the pair
$(\hat{K},\hat{\alpha})$ that jointly maximise \eqref{eq:density_ml}
suggests that $\hat{K}$ serves as a good number of bins for a regular
histogram of the observed data.



\section{Alternative strategies}\label{sec:other_methods}
To calibrate the performance of the proposed method, some variations
of the strategy for selecting sample sizes are considered. This
section considers some alternative measures of the Monte Carlo error
of a sampler, to be used in place of the divergence estimates
$\hat{e}^{(j)}_{n^{(j)}}$ or $\hat{\delta}^{(j)}_{n^{(j)}}$ in the
algorithm of Section \ref{sec:alg}.
\subsection{Chi-squared goodness of fit statistic}\label{sec:fox}
In the context of particle filters, \cite{Fox03} proposed a method for
choosing the number of samples $n$ required from a single sampler to
guarantee that, under a chi square approximation, with a desired
probability $(1-\delta)$ the Kullback-Leibler divergence between the
binned empirical and true distributions does not exceed a certain
threshold $\varepsilon$. This was achieved by noting an identity
between $2n$ times this divergence and the likelihood ratio statistic
for testing the true distribution against the empirical distribution,
assuming the true distribution had the same number of bins, $K$, as
the observed empirical distribution. Since the likelihood ratio
statistic should approximately follow a chi-squared distribution with
$K-1$ degrees of freedom, this suggested a sample size of
\begin{equation}
n=\chi^2_{K-1,1-\delta}/(2\varepsilon) \label{eq:chi_n},
\end{equation}
where $\chi^2_{K-1,1-\delta}$ is the $1-\delta$ quantile of that
distribution.

Adapting this idea to the algorithm of Section \ref{sec:alg} simply
requires a rearrangement of \eqref{eq:chi_n} to give the
approximate error as a function of sample size,
\begin{equation}
\varepsilon=\chi^2_{K-1,1-\delta}/(2n) \label{eq:chi_e}.
\end{equation}
This error estimate can be substituted directly into the algorithm in
place of the Monte Carlo divergence error estimate
$\hat{e}^{(j)}_{n^{(j)}}$ to provide an alternative scheme for
choosing sample sizes when using loss function
$\mathcal{L_{\mathrm{max}}}$. The same (arbitrary) value of $\delta$
must be used for each rival sampler, and here this was specified as
$\delta=0.05$ although the results are robust to different choices.

By the central limit theorem, the chi-squared distribution quantiles
grow near-linearly with the degrees of freedom parameter for $K> 30$
\citep{fisher1959}, so it should be noted that \eqref{eq:chi_e}, which
depends only on the number of bins, has much similarity, and almost
equivalence, with the Miller-Madow estimate of entropy error cited in
Section \ref{sec:mc_divergence_estimation}. By the reasoning given in
Section \ref{sec:mc_divergence_estimation}, use of this error function
should show some similarity in performance with the proposed method,
but be less robust to distinguishing differences in distributions
beyond the number of non-empty bins.

Recall from Section \ref{sec:sequential} that the sequential
allocation strategy for minimising the loss function
$\mathcal{L_{\mathrm{ave}}}$ requires an estimate of the expected
reduction in error which would be achieved from obtaining another
observation from a sampler. Since this error criterion depends
entirely upon the number of non-empty bins $K$, in this case an
estimate is required for the probability of the new observation
falling into a new bin.
A simple empirical estimate of the probability of falling into a new
bin is provided by the proportion of samples after the first one that
have fallen into new bins, given by $(K-1)/(\ns-1)$. Note that this
estimate will naturally carry positive bias, since discovery of new
bins should decrease over time, and so a sliding window of this
quantity might be more appropriate in some contexts.

\subsection{Reference points}\label{sec:foi}
As a convergence diagnostic for transdimensional samplers,
\cite{sisson_fan07} proposed running replicate sampling chains for the
same target distribution, and comparing the variability across the
chains of the empirical distributions of a distance-based function of
interest. The method requires that the target $\pi$
be a probability distribution for a point process, 
and maps multidimensional sampled tuples of \textit{events} from $\pi$
to a fixed-dimension space. 
Specifically, a set of \textit{reference points}
$\mathcal{V}$ are chosen, and for any sampled tuple of events the
distance from each reference point to the closest event in the tuple
is calculated. Thus $\pi$ is summarised by a
$|\mathcal{V}|$-dimensional distribution, where $|\mathcal{V}|$ is the
number of reference points in $\mathcal{V}$.

One example considered in \cite{sisson_fan07} is a Bayesian
continuous-time changepoint analysis of a changing regression model
with an unknown number of changepoint locations. A variation of this
example is analysed in Section \ref{sec:results_multivariate} in this
article, where instead the analysis will be for the canonical problem
of detecting changes in the piecewise constant intensity function
$\lambda(t), t\in[0,1]$ of an inhomogeneous Poisson process
\citep[see][and the subsequent literature]{raftery_akman1986}. For
Poisson process data, there are two natural functions of interest
which could be evaluated at each reference point. The first is the
distance to the nearest changepoint, the second is the intensity
level. Both will be considered in Section
\ref{sec:results_multivariate}. Note that in \cite{sisson_fan07},
reference points are selected from random components from an initial
sample from the single target distribution. Here, since there are
multiple target distributions, a grid of one hundred uniformly spaced
points across the domain $[0,1]$ is used.

The convergence diagnostic of \cite{sisson_fan07} did not formally
provide a method for calibrating error or selecting sample size. Here,
to compare the performance of the proposed sample size algorithm of
Section \ref{sec:alg}, 
the sum across the reference points of the
Monte Carlo variances of either of these functions of interest is used
as the error criterion in the algorithm.

\subsection{\textit{Ad hoc} strategies}\label{sec:ad_hoc_strategies}
To demonstrate the value of the sophisticated sample
size selection strategies given above, two simple strategies which have
similar motivation but are otherwise \textit{ad hoc} are included
in the numerical comparisons of Section \ref{sec:examples}. These
strategies are now briefly explained.

\subsubsection{Extent} \label{sec:extent}
The \textit{extent} of a distribution is the exponential of its
entropy, and was introduced as a measure of spread by
\cite{campbell66}. A simple strategy might be to choose sample size
proportional to the estimated squared extent of $\pi$,
$\exp\{2H(\hat{\pi})\}$. Note that the Gaussian distribution
$\mathrm{N}(\mu,\sigma^2)$, has an extent which is directly
proportional to the standard deviation $\sigma$, and so in the
univariate Gaussian example which will be considered in Section
\ref{sec:results_univariate}, this sample allocation strategy will be
approximately equivalent to the optimal strategy when minimising the
maximum Monte Carlo error of the sample means (\textit{cf.} Section
\ref{sec:loss}).

\subsubsection{Jensen-Shannon divergence}\label{sec:jsd}
\cite{Robert2005} present a class of convergence tests for monitoring
the stationarity of the output of a sampler from a single run which
operate by splitting the current sample $(x_1,x_2,\ldots,x_n)$ in two
and quantifying the difference between the empirical distributions of
the first half of the sample $(x_1,x_2,\ldots,x_{\lfloor
  n/2\rfloor})$, and the second half of the sample $(x_{\lfloor
  n/2\rfloor+1},x_{\lfloor n/2\rfloor+2},\ldots,x_{n})$. For
univariate samplers the Kolmogorov-Smirnov test, for example, is used
to obtain a p-value as a measure of evidence that the second half of
the sample is different from the first, and hence neither half is
adequately representative of the target. The test statistics which are
used condition on the sample size, and so the sole purpose of these
procedures is to investigate how well the sampler is mixing and
exploring the target distribution.

To adapt these ideas to the current context, any mixing issues can
first be discounted by splitting the sample for each target in half
by allocating the samples into two groups alternately, so that
the distribution of, say, $(x_1,x_3,\ldots,x_{n-1})$ can be compared
with the distribution of $(x_2,x_4,\ldots,x_n)$. This method of
splitting up the sample is also computationally much simpler in a
streaming context, as incrementing the sample size $n$ does not
change the required groupings of the existing samples. Let
$\hat{\pi}_n^{\mathrm{odd}}$ and $\hat{\pi}_n^{\mathrm{even}}$ be the
respective empirical distributions of these two subsamples.

A crude variation on using the Monte Carlo divergence error criteria of
\eqref{eq:ehat_KL} is to estimate the error of the sampler by
the Jensen-Shannon divergence of $\hat{\pi}_n^{\mathrm{odd}}$ and
$\hat{\pi}_n^{\mathrm{even}}$,
\begin{equation}
\hat{e}_{JSD, \ns}=\jsd(\hat{\pi}_n^{\mathrm{odd}},\hat{\pi}_n^{\mathrm{even}}) = H(\hat{\pi}_n)-\frac{H(\hat{\pi}_n^{\mathrm{odd}}) + H(\hat{\pi}_n^{\mathrm{even}})}{2}.\label{eq:ehat_jsd}
\end{equation}
If sufficiently many samples have been taken for $\hat{\pi}_n$
to be a good representation of the target distribution, then both halves
of the sample should also provide reasonable approximations of the
target and therefore have low divergence between one another.

As in Section \ref{sec:efficient_calculation}, calculation of
\eqref{eq:ehat_jsd} during sampling can be updated at each iteration
very quickly. Let $i'$ be the bin in which the $n$th observation
falls. Then, for example, updating the first term of
\eqref{eq:ehat_jsd} simply requires
\begin{equation}
H(\hat{\pi}_n) = H(\hat{\pi}_{n-1})+\log(n)-\log(n-1)-n_{i'}\log(n_{i'})+(n_{i'}-1)\log(n_{i'}-1).
\end{equation}

\section{Examples}\label{sec:examples}
The methodology from this article is demonstrated on three different data
problems. The first two examples assume only two or three data
processes respectively, to allow a detailed examination of how
the allocation strategies differ. Then finally a larger scale example
with 400 data processes is considered, derived from the IEEE VAST 2008
Challenge concerning communication network anomaly detection.

\subsection{Small scale examples}\label{sec:small_scale_examples}

Two straightforward, synthetic examples are now considered. The first
is a univariate problem of fixed dimension with two Gaussian target
distributions, and the second is a transdimensional problem of
unbounded dimension, concerning the changepoints in the piecewise
constant intensity functions of three inhomogeneous Poisson
processes. In both examples, it is assumed that \textit{a priori}
nothing is known about the target distributions and that computational
limitations determine that only a fixed total number of samples can be
obtained from them overall, which will correspond to an average of $50{,}000$
samples per target distribution.

Both loss functions from Section \ref{sec:loss} are considered,
measuring either the maximum error or average error across the target
samplers. For each loss function, the following sample size allocation
strategies are considered:
\begin{enumerate}
\item ``Fixed'' --- The default strategy, $50{,}000$ samples are
  obtained from each sampler.
\item Dynamically, aiming to minimise the expected loss, with sampling
  error estimated using the following methods:
\begin{enumerate}
\item ``Grassberger'' --- Monte Carlo divergence error
  estimation from Section \ref{sec:mc_divergence_estimation};
\item ``Fox'' --- the $\chi^2$ goodness of fit statistic of Section
  \ref{sec:fox};
\item ``Sisson'' (only for the transdimensional example) --- the Monte
  Carlo variances of one of the two candidate fixed dimension
  functions from Section \ref{sec:foi} evaluated at 100 equally spaced
  reference points (denoted ``Sisson-i'', for the intensity function,
  ``Sisson-n'' for the distance to nearest changepoint function);
\item ``Extent'' and ``JSD'' --- two \textit{ad hoc} criteria from
  Section \ref{sec:ad_hoc_strategies}.
\end{enumerate}
\end{enumerate}
Each sample size allocation strategy is evaluated over a large number
of replications $M$, where $M=1 \mathrm{~million}$ or $M=100{,}000$
respectively in the two examples.

Good performance of a sample allocation strategy is measured by the
chosen loss function when applied to the realised Monte Carlo
divergence error $e_{\KL}$ for each sampler. Good estimates of the
true values of $e_{\KL}$ are obtained by calculating the
Jensen-Shannon divergence of the Monte Carlo empirical distributions
obtained from the $M$ runs (\textit{cf.} Section
\ref{sec:mc_divergence}).

Note that in all simulations, the same random number generating seeds
are used for all strategies, so that all strategies are making
decisions based on exactly the same samples.


\subsubsection{Univariate target distributions}\label{sec:results_univariate}

In the first example, a total of 100{,}000 samples are drawn from two
Gaussian distributions, where one Gaussian has twice the standard
deviation of the other:
$
\pi_1(x)=\mathrm{N}(x|0,1)$, $
\pi_2(x)=\mathrm{N}(x|0,4).
$
Note that if these two distributions were considered on different
scales they would be equivalent; but when losses in estimating the
distributions are measured on the same scale, then they are not
equivalent. For discretising the distributions, the following bins
were used: $(-\infty,-10)$, $[-10,-9.8)$, $[-9.8,-9.6)$, $\ldots$, $[9.6,9.8)$, $[9.8,10)$, $[10,\infty)$. This corresponds to an interior range of plus or
minus five times the largest of the standard deviations of the two
targets, divided into 100 evenly spaced bins, along with two extra
bins for the extreme tails.
Results are robust to allowing wider ranges or more bins, but are
omitted from presentation. For further validation, a simple experiment
was conducted using the method from Section
\ref{sec:bayesian_bin_width} on $[-10,10]$: 100{,}000 samples were
simulated from each of $\pi_1$ and $\pi_2$, leading to estimates
$\hat{K}=92$ and $\hat{K}=76$ respectively, suggesting 100 as a good
number of bins for fitting these densities.

\begin{figure}[ht!]
\centering
\makebox{\includegraphics[width=.49\textwidth]{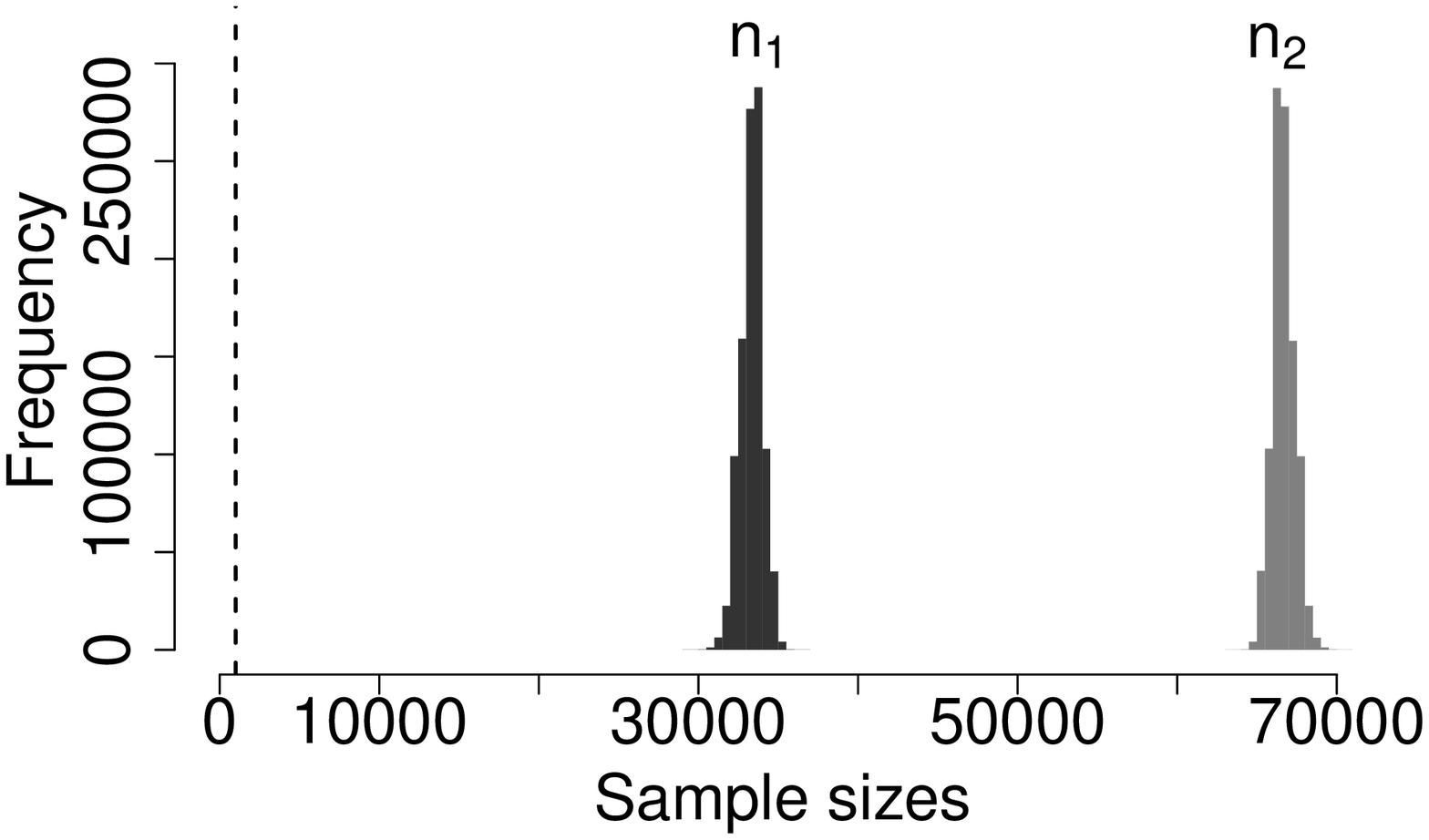}}
\makebox{\includegraphics[width=.49\textwidth]{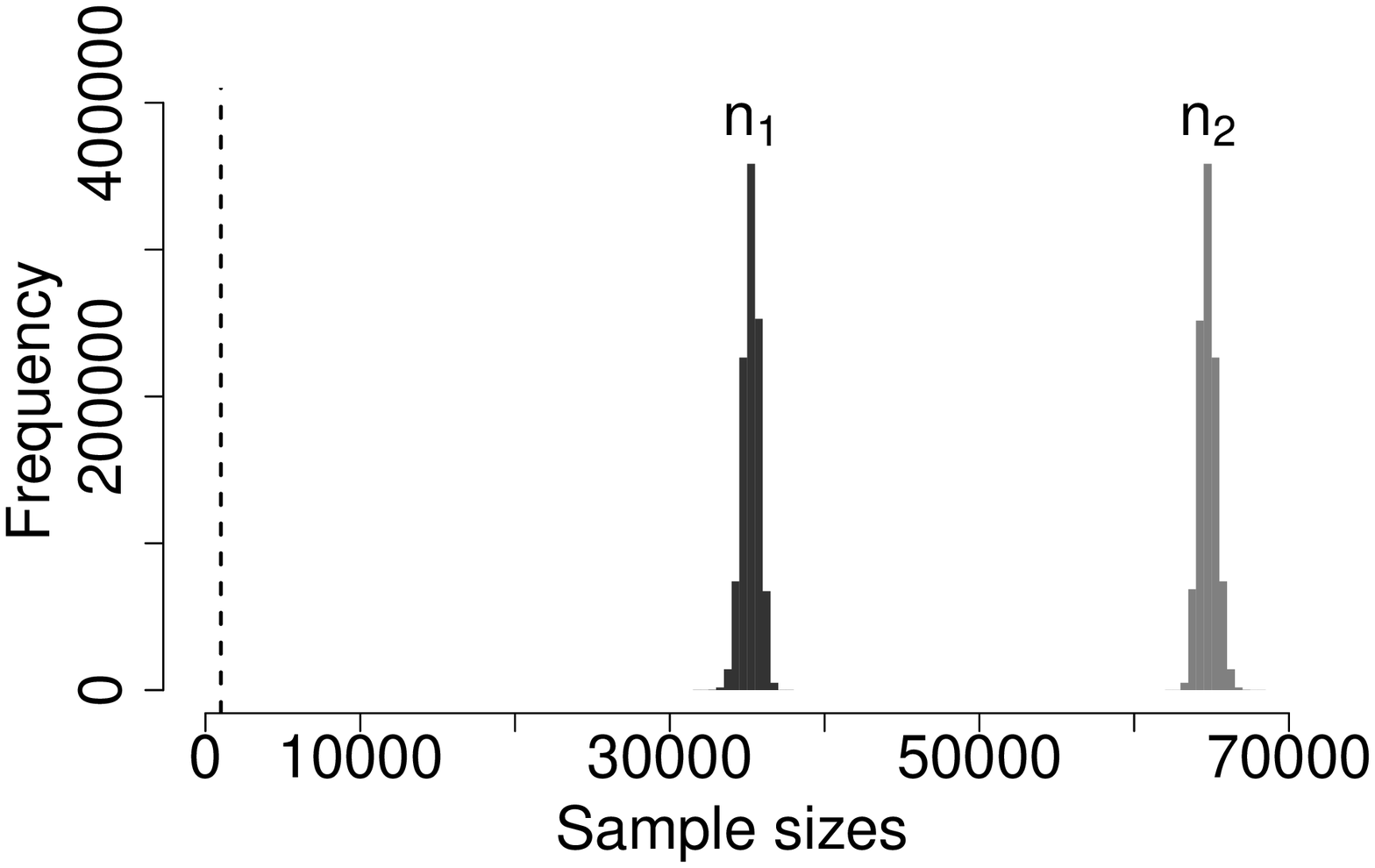}}\\
\makebox{\includegraphics[width=.49\textwidth]{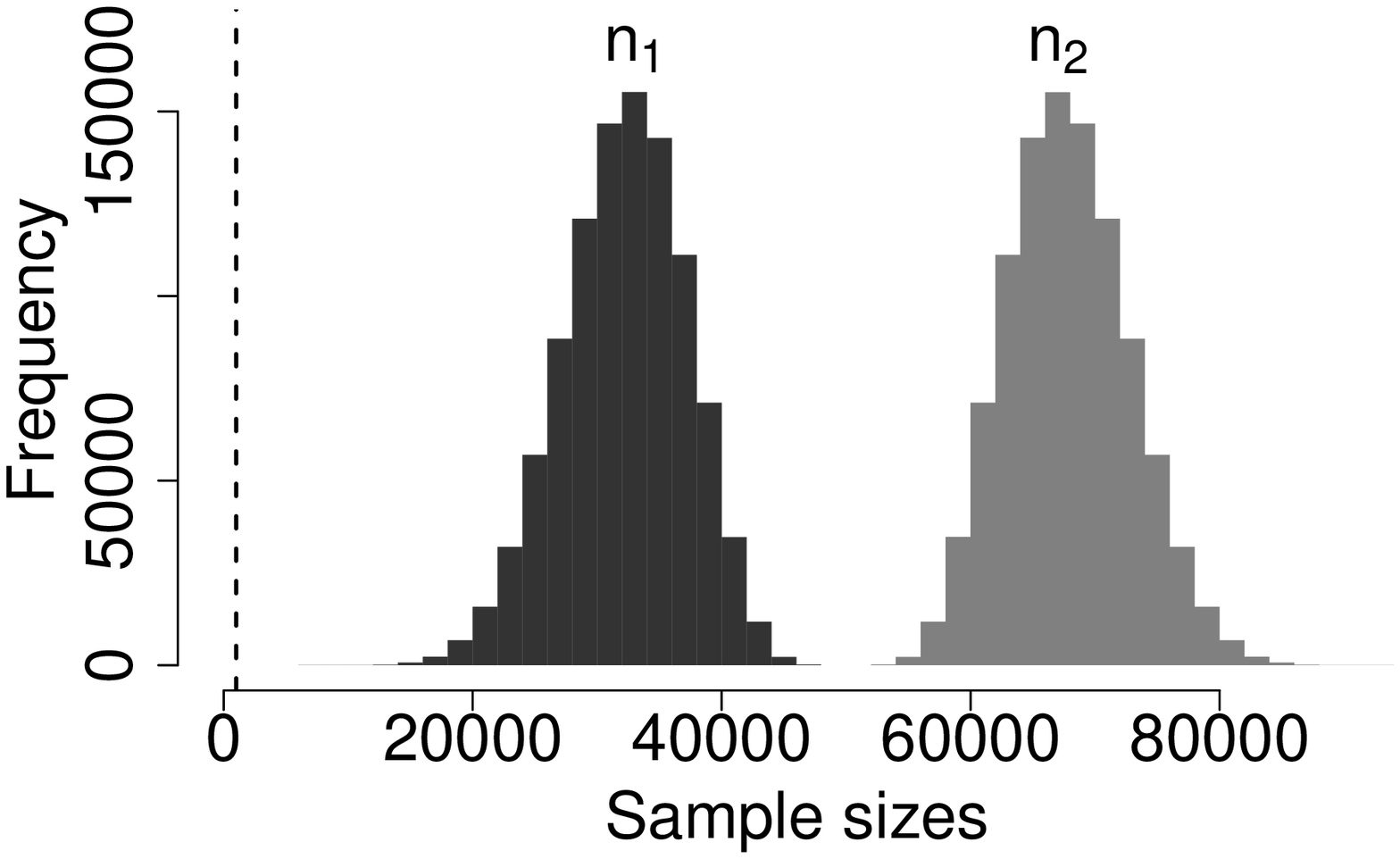}}
\makebox{\includegraphics[width=.49\textwidth]{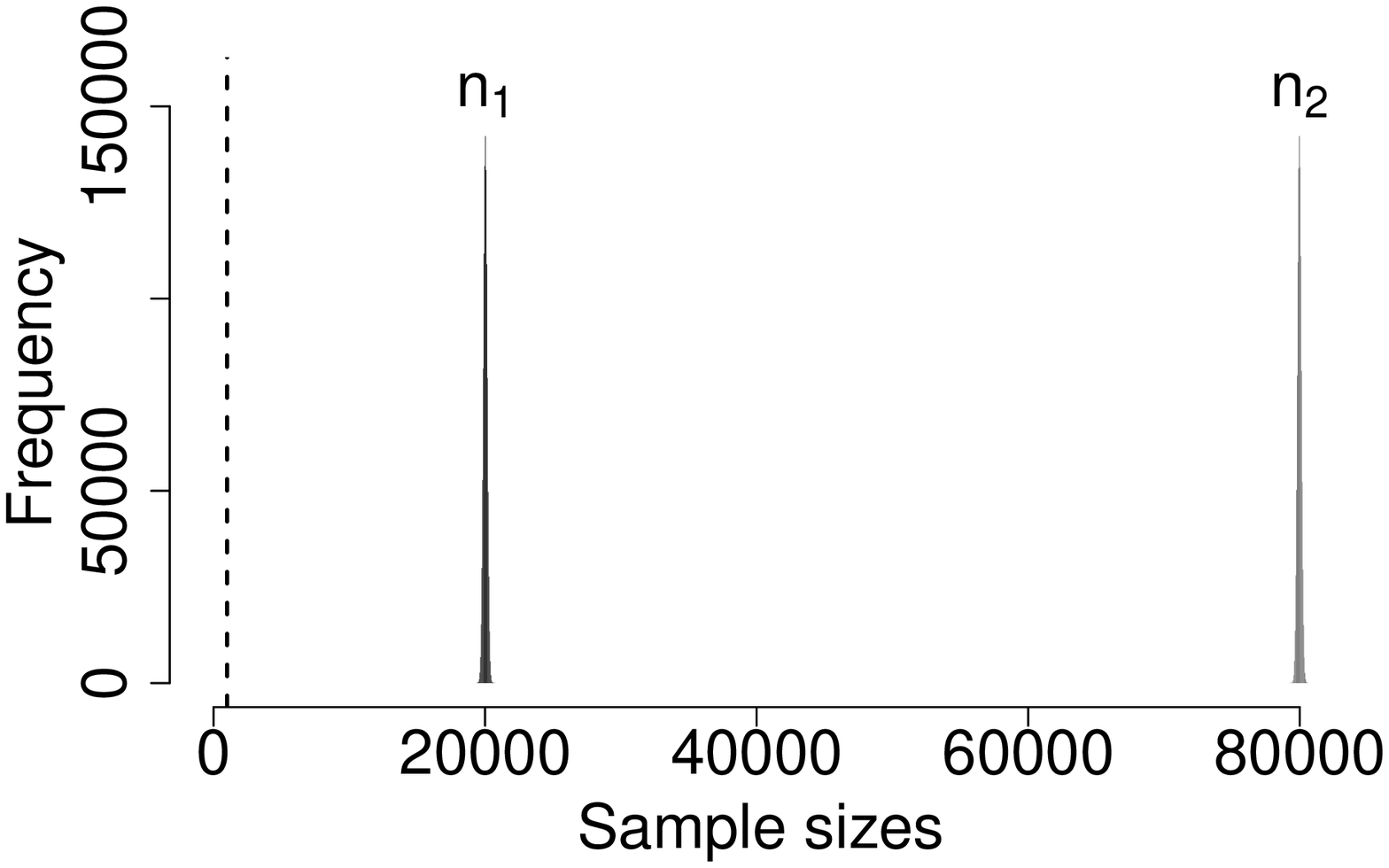}}\\
\caption{\label{fig:univariate_sample_sizes}Distributions 
of the sample sizes $(n_1,n_2)$ allocated to the two rival univariate
samplers under the loss function $\mathcal{L_{\mathrm{max}}}$ when
constrained to a total of $n_1+n_2=100{,}000$ samples, using the
allocation strategies ``Grassberger'' (top left), ``Fox'' (top right),
``JS'' (bottom left), ``Extent'' (bottom right). }
\end{figure}

The varying sample sizes obtained from each target from the 1 million
simulations using each of the sample allocation strategies listed
above and the loss function $\mathcal{L_{\mathrm{max}}}$ are shown in
Fig.  \ref{fig:univariate_sample_sizes}. Tables
\ref{tab:univariate_losses_max} and \ref{tab:univariate_losses_ave}
show the mean sample sizes and the implied Monte Carlo divergence error for
each target distribution using each of the sample allocation
strategies listed above; the two tables correspond to the two choices
of loss function for combining errors.

Table \ref{tab:univariate_losses_max} gives the results under the loss
function \eqref{eq:max_loss} which calculates the maximal error across
samplers. Optimal performance would imply approximately equal Monte
Carlo divergence errors for the two targets, and the proposed strategy
based on Grassberger's entropy bias estimate is by far the closest to
achieving this objective. Interestingly, note that under this best
strategy, the average sample sizes are almost exactly in the ratio
1:2, the same ratio as the true standard deviations of the target
distributions. Recall from Section \ref{sec:loss}, that such a ratio
is optimal in another sense, minimising the \textit{average} Monte
Carlo errors of the sample means. This contrast highlights the
importance of carefully specifying the desired error criterion as well
as the correct loss function.

One of the two \textit{ad hoc} strategies based on calculating the
Jensen-Shannon divergence between the two halves of the sample is only
slightly outperformed by the $\chi^2$ goodness of fit method; however,
note in Figure \ref{fig:univariate_sample_sizes} the much higher
variance of the sample sizes with the JSD method, which is indicative
of an unreliable strategy. The other \textit{ad hoc} method which
takes sample sizes proportional to the extent of the empirical
distributions is seen to overcompensate for the higher variance of the
second Gaussian, and performs worse than even the default equal sample
size strategy. For that strategy, note the sample sizes are almost
exactly in the ratio 1:4, the same ratio as the true variances of the
target distributions. Recall from Section \ref{sec:extent} that such a
strategy is approximately equivalent to minimising the
\textit{maximum} Monte Carlo errors of the sample means, which was
noted in Section \ref{sec:loss} to imply such an allocation ratio.

\begin{table}
\caption{\label{tab:univariate_losses_max}Monte Carlo divergence error
  $e_{\KL}$ ($\times 10^{-4}$) of univariate target distribution
  estimates for each allocation strategy under
  $\mathcal{L_{\mathrm{max}}}$. Average sample size are in parentheses.}
\centering \fbox{%
\begin{tabular}{*{6}{c}}
& Equal & Grassberger & Fox & Extent & JSD \\
\hline
\multirow{2}{*}{$\pi_1$}& 4.629 & 6.86864 & 6.50219 &11.3688 &7.31157\\
& (50{,}000) & (33{,}338) & (35{,}229) & (20{,}022) & (32{,}159)\\
\multirow{2}{*}{$\pi_2$}& 9.03931  & 6.87318  & 7.06507 &5.77444 &6.79011\\
& (50{,}000) & (66{,}662) & (64{,}771) & (79{,}978) & (67{,}841) \\
\hline
$\mathcal{L_{\mathrm{max}}}$& 9.03931 & \textbf{6.87318} & 7.06507 & 11.3688 &7.31157\
\end{tabular}}
\end{table}

Table \ref{tab:univariate_losses_ave} gives the results under the loss
function $\mathcal{L_{\mathrm{ave}}}$ \eqref{eq:ave_loss} which
calculates the average error across samplers. The Monte Carlo
divergence strategy based on Grassberger's entropy bias estimate
performs best, although the the $\chi^2$ goodness of fit method also
performs very well here. The contrasting sample sizes between the loss
functions $\mathcal{L_{\mathrm{max}}}$ and
$\mathcal{L_{\mathrm{ave}}}$ for all dynamic allocation strategies are
noteworthy, as remarked in Section \ref{sec:loss}.

\begin{table}
\caption{\label{tab:univariate_losses_ave}Monte Carlo divergence error
  $e_{\KL}$ ($\times 10^{-4}$) of univariate target distribution
  estimates for each allocation strategy under 
  $\mathcal{L_{\mathrm{ave}}}$.
}
\centering \fbox{%
\begin{tabular}{*{6}{c}}
& Equal & Grassberger & Fox & Extent & JSD \\
\hline
\multirow{2}{*}{$\pi_1$}& 4.629 & 5.51872 & 5.42576 &6.85805 & 5.50771\\
& (50{,}000) & (41{,}670) & (42{,}398) & (33{,}361) & (41{,}863)\\
\multirow{2}{*}{$\pi_2$}& 9.03931 & 7.80652  & 7.90033 &6.8733 & 7.8400\\
& (50{,}000) & (58{,}330) & (57{,}602) & (66{,}639) & (58{,}137) \\
\hline
$\mathcal{L_{\mathrm{ave}}}$& 6.83416 & \textbf{6.66262} & 6.66304 & 6.86568 & 6.67385\\
\end{tabular}}
\end{table}

\subsubsection{Transdimensional mixture target
  distributions}\label{sec:results_multivariate}

For a more complex example, simulated data were generated from three
inhomogeneous Poisson processes on $[0,1]$ with different piecewise
constant intensity functions. In each case, prior beliefs for the
intensity functions were specified by a homogeneous Poisson process
prior distribution on the number and locations of the changepoints and
independent, conjugate gamma priors on the intensity levels. The three
rival target distributions for inference are the Bayesian marginal
posterior distributions on the number and locations of the
changepoints for each of the three processes.

\begin{figure}[ht!]
\centering
\makebox{\includegraphics[width=\textwidth]{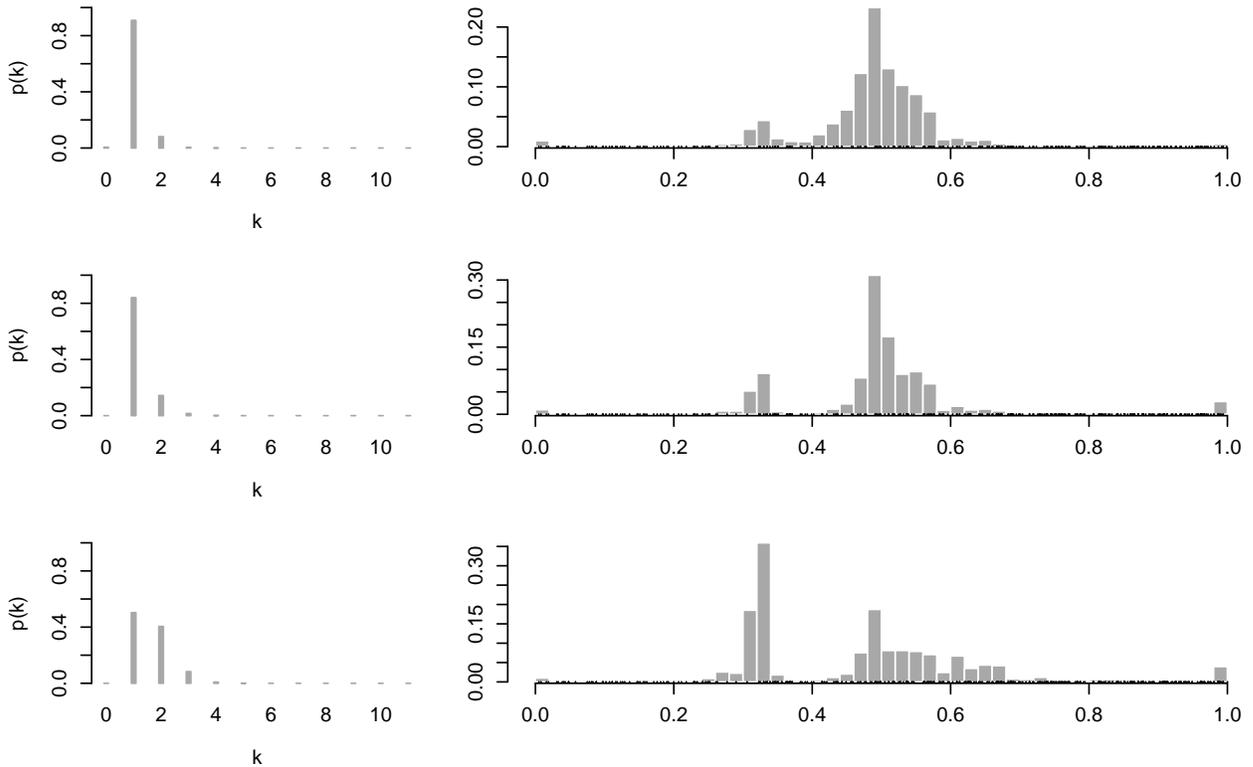}}
\caption{\label{fig:targets} Monte Carlo estimates of the target
  posterior changepoint distributions for the three simulated Poisson
  processes. The rows are the three target distributions
  $\pi_1,\pi_2,\pi_3$; the columns show the posterior distribution of
  the number of changepoints $k$ and a binned one-dimensional projection
  of the target where each bar shows the probability of a changepoint
  falling in that bin.}
\end{figure}

\begin{figure}[ht!]
\centering
\makebox{\includegraphics[width=\textwidth]{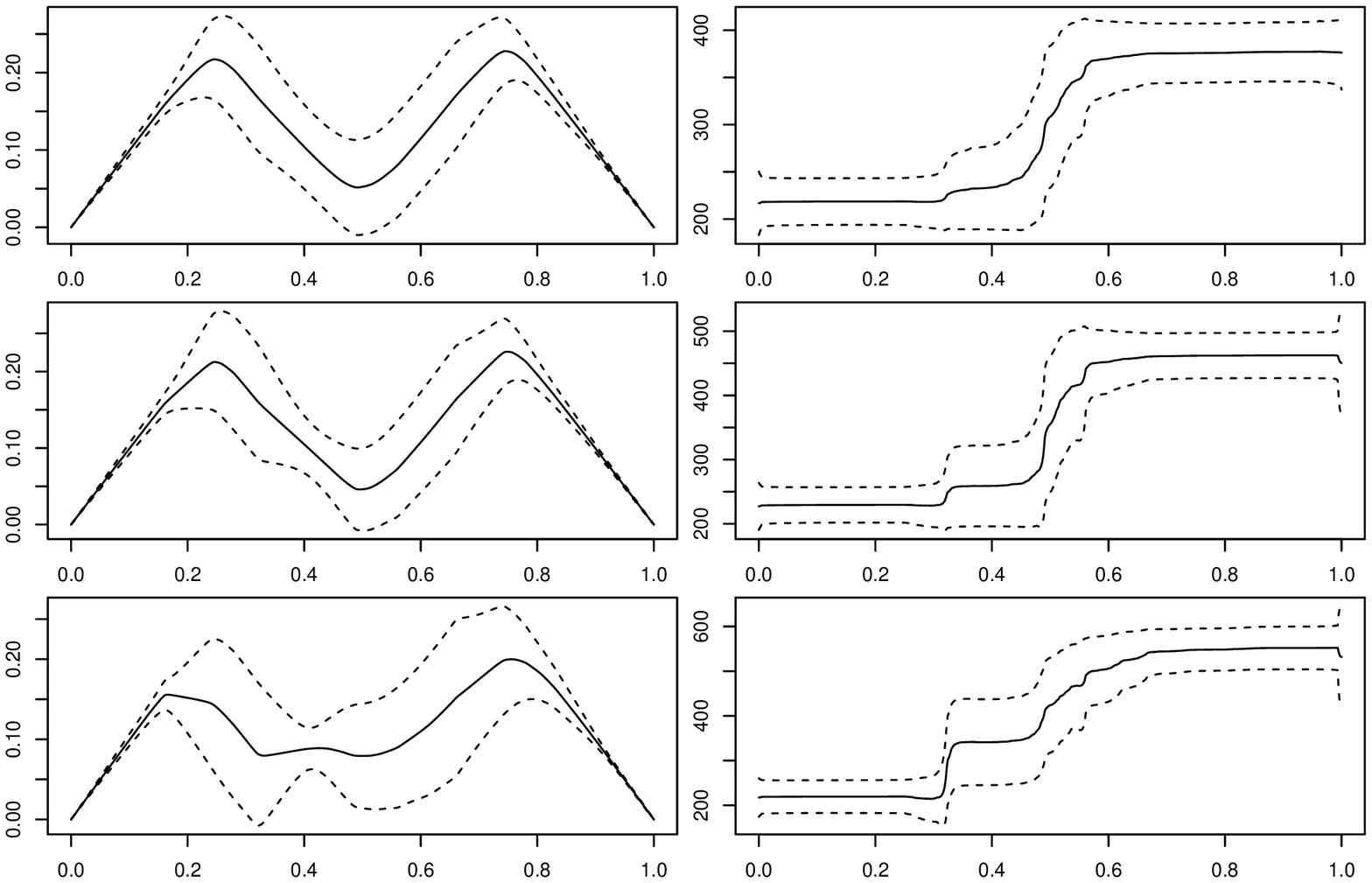}}
\caption{\label{fig:fois} Monte Carlo estimates of two expectations
  with respect to the changepoint target posterior distributions.
  The rows are the three target distributions
  $\pi_1,\pi_2,\pi_3$. Left: the expected distance to the nearest
  changepoint. Right: The intensity function of the data process. The
  solids lines are the posterior means, and the dotted lines indicate
  one standard deviation.}
\end{figure}

Each of the three simulated Poisson processes had two changepoints,
located at $1/3$ and $2/3$. The intensity levels of the three
processes were respectively: $(200,300,400)$, $(200,350,500)$,
$(200,400,600)$, so the processes differed only through magnitudes of
intensity changes. To make the target distributions closer and
therefore make the inferential problem harder, in each case the prior
expectation for the number of changepoints was set to 1. For
illustration of the differences in complexity of the resulting
posterior distributions for the changepoints of the three processes,
large sample estimates of the true, discretised posterior
distributions are shown in Fig. \ref{fig:targets}, based upon one
trillion reversible jump Markov chain Monte Carlo samples
\citep{green95}. Note that the different target distributions place
different levels of mass on the number of changepoints, and therefore
on the dimension of the problem. In all cases there is insufficient
information to strongly detect both changepoints, and so much of the
mass of the posterior distributions is localised at a single
changepoint at $1/2$, the midpoint of the two true
changepoints. Additionally, Fig. \ref{fig:fois} shows the posterior
variance of two functions of interest identified in Section
\ref{sec:foi} for $t\in[0,1]$: the distance to the nearest
changepoint, and the intensity level.

To determine sample sizes, reversible jump Markov chain Monte Carlo
simulation was used to sample from the marginal posterior
distributions of the changepoints, and the chains were thinned with
only one in every fifty iterations retained to give approximately
independent samples. To discretise the distributions, the interval
$[0,1]$ was divided into 50 equally sized bins; while for a single
dimension this would be fewer bins than were used in the previous
section, here the bins are applied to each dimension of a mixture
model of unbounded dimension, meaning that actually a very large
number of bins are visited; computational storage issues can begin to
arise when using an even larger number of bins, simply through storing
the frequency counts of the samples.

\begin{figure}[ht!]
\centering
\makebox{\includegraphics[width=.49\textwidth]{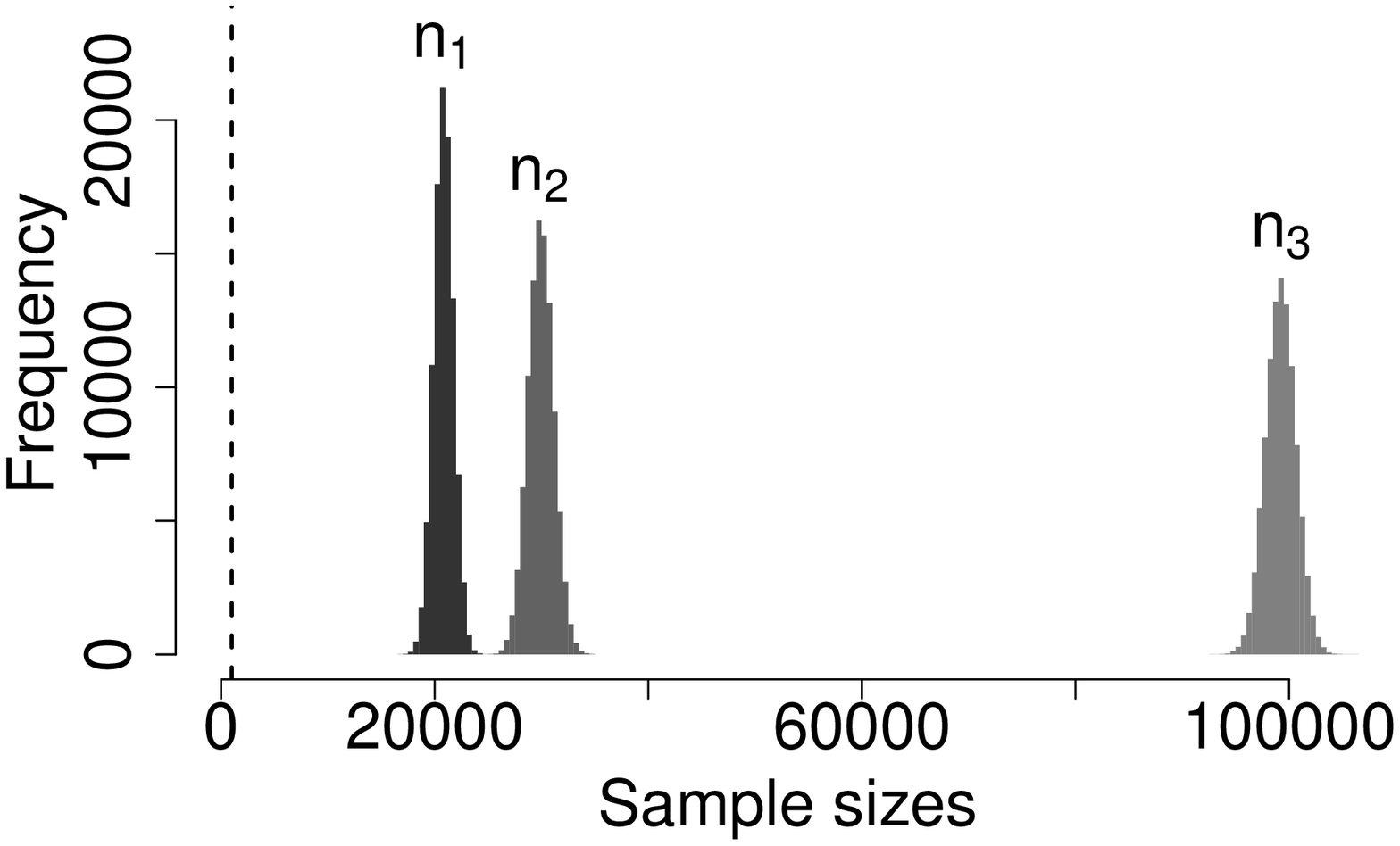}}
\makebox{\includegraphics[width=.49\textwidth]{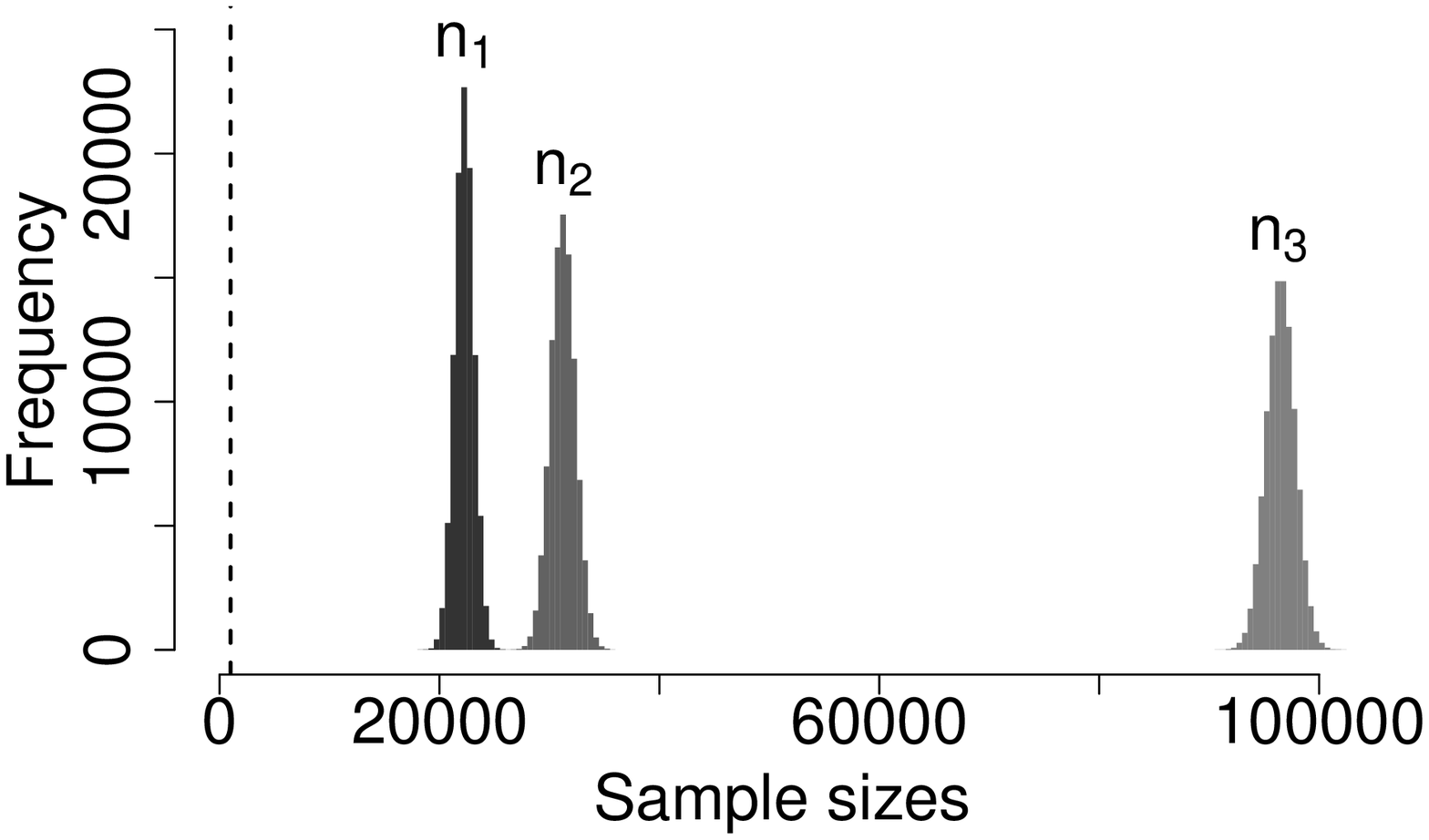}}\\
\makebox{\includegraphics[width=.49\textwidth]{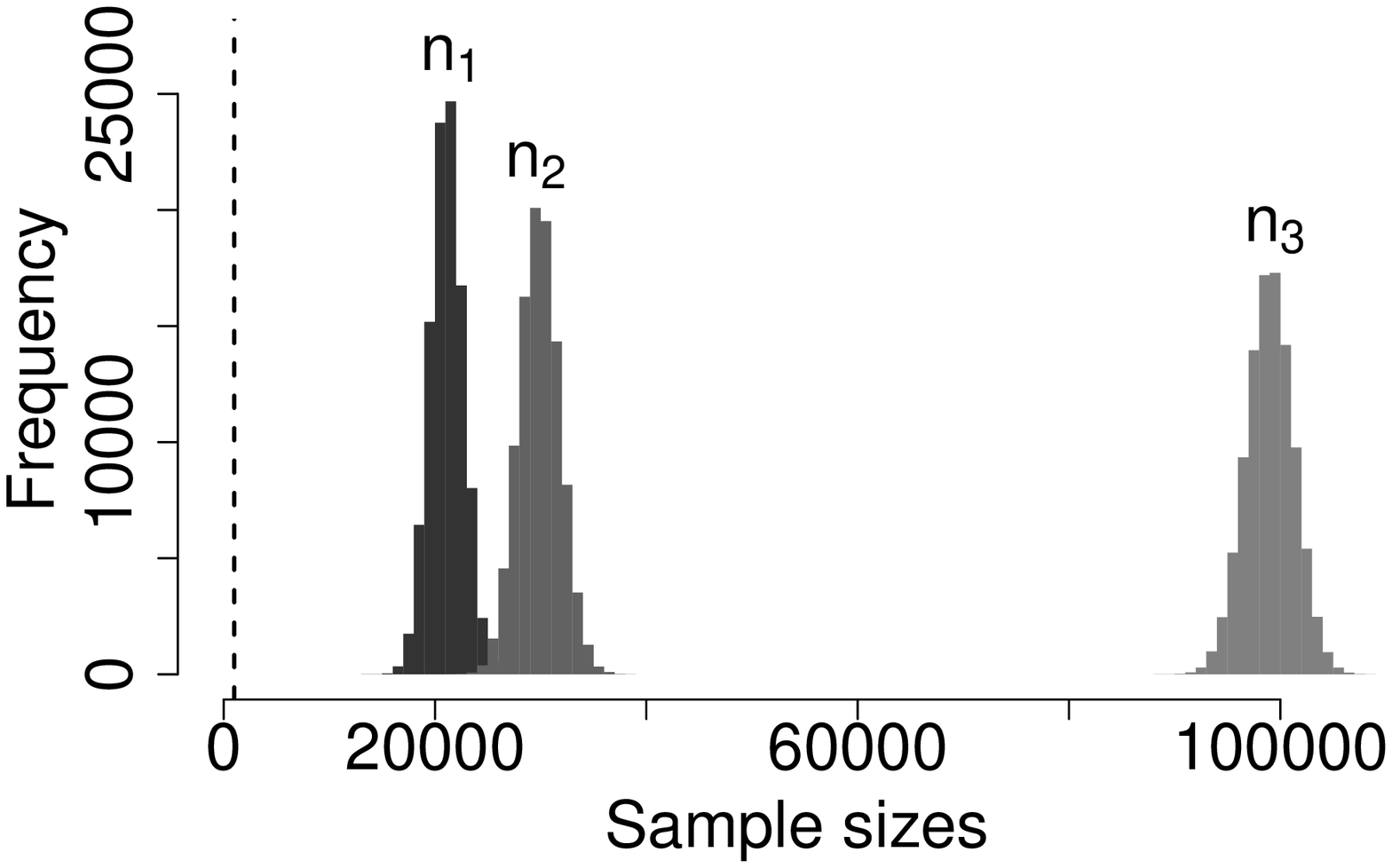}}
\makebox{\includegraphics[width=.49\textwidth]{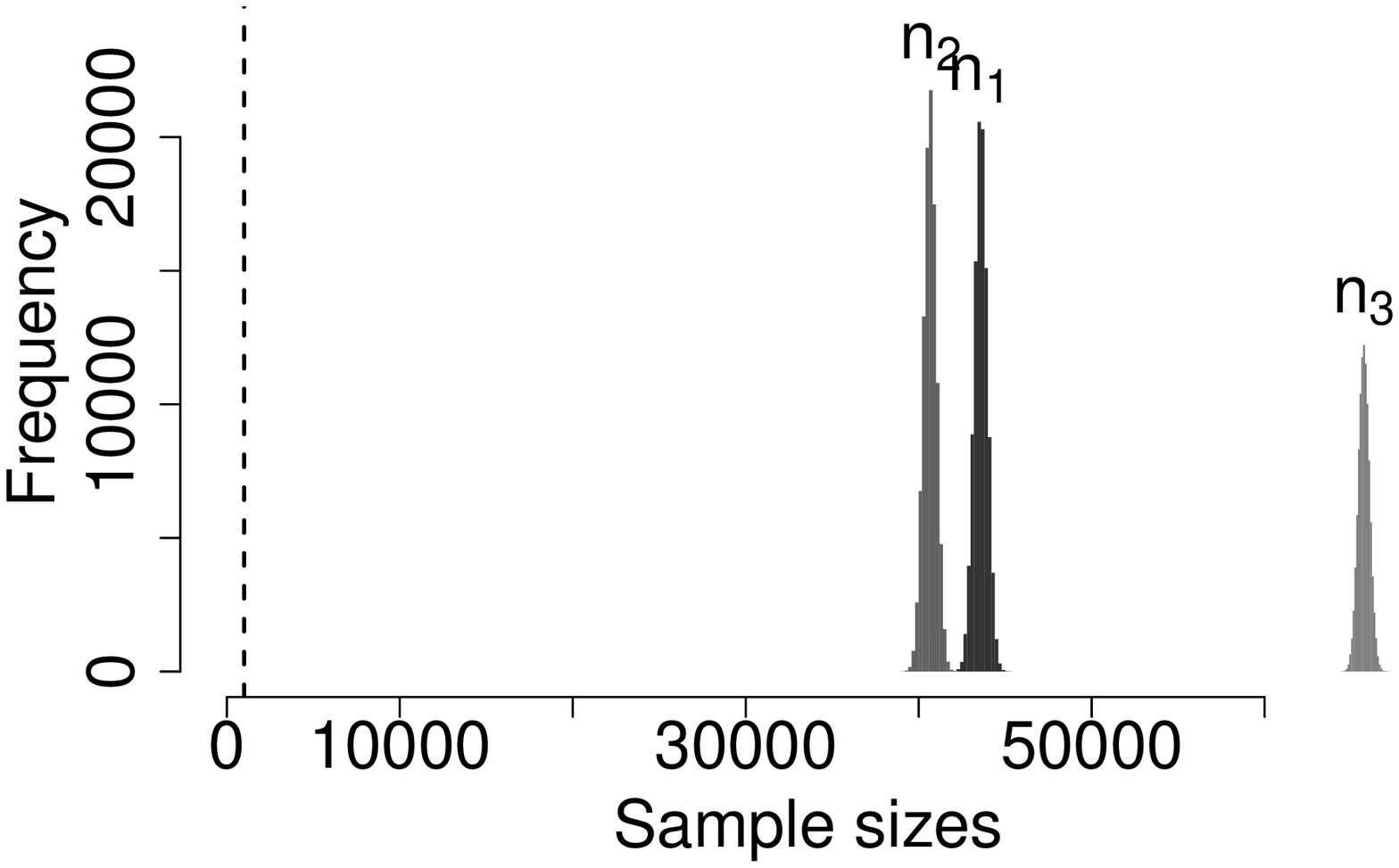}}\\
\caption{\label{fig:3target_sample_sizes} Distributions 
of the sample sizes $(n_1,n_2,n_3)$ allocated to the three rival
transdimensional samplers under the loss function
$\mathcal{L_{\mathrm{max}}}$ when constrained to a total of
$n_1+n_2+n_3=150{,}000$ samples, using the allocation strategies
``Grassberger'' (top left), ``Fox'' (top right), ``JS'' (bottom left),
``Sisson-n'' (bottom right). }
\end{figure}

Fig. \ref{fig:3target_sample_sizes} shows the distributions of sample
sizes obtained from a selection of the strategies over $M=100{,}000$
repetitions, and Tables \ref{tab:transdim_losses_max} and
\ref{tab:transdim_losses_ave} show results from the different
strategies examined for these more complex transdimensional
samplers. Performance is similar to the previous section, with the
Grassberger entropy bias correction method performing best.

\begin{table}
\caption{\label{tab:transdim_losses_max}Monte Carlo divergence error
  $e_{\KL}$ ($\times 10^{-2}$) of transdimensional sampler target distribution
  estimates for each allocation strategy under 
  $\mathcal{L_{\mathrm{max}}}$. Average sample sizes are in parentheses.}
\centering \fbox{%
\begin{tabular}{*{8}{c}}
& Equal & Grassberger & Fox & Extent & JSD & Sisson-i & Sisson-n \\
\hline
\multirow{2}{*}{$\pi_1$}& 2.69336 & 4.26407 & 4.11478 & 6.85586 & 4.24113 & 3.69798 & 2.88843\\
& (50{,}000) & (20{,}838) & (22{,}257) & (8{,}725) & (21{,}105) & (27{,}180)& (43{,}595) \\
\multirow{2}{*}{$\pi_2$}& 3.74377  & 4.79495  & 4.69796 & 8.43441 & 4.80385 & 3.97563 & 4.13582\\
& (50{,}000) & (29{,}936) & (31{,}225) &  (9{,}510) & (29{,}877) & (44{,}130) & (40{,}660) \\
\multirow{2}{*}{$\pi_3$}& 6.77106 & 4.85438  & 4.92019 & 4.22777 & 4.85987 & 5.43455 & 5.93004\\
& (50{,}000) & (99{,}226) & (96{,}518) &  (131{,}765) & (99{,}018) & (78{,}690) & (65{,}745) \\
\hline
$\mathcal{L_{\mathrm{max}}}$& 6.77106 & \textbf{4.85438} & 4.92019 & 6.85586 & 4.85961 & 5.43455 & 5.93004\\
\end{tabular}
}
\end{table}

\begin{table}
\caption{\label{tab:transdim_losses_ave}Monte Carlo divergence error
  $e_{\KL}$ ($\times 10^{-2}$) of transdimensional sampler target distribution
  estimates for each allocation strategy under
  $\mathcal{L_{\mathrm{ave}}}$. 
}
\centering \fbox{%
\begin{tabular}{*{8}{c}}
& Equal & Grassberger & Fox & Extent & JSD & Sisson-i & Sisson-n \\
\hline
\multirow{2}{*}{$\pi_1$}& 2.69336 & 3.06906 & 3.05301 & 3.79343 & 3.06637 & 3.11194 & 2.78047\\
& (50{,}000) & (38{,}734) & (39{,}129) & (25{,}910) & (38{,}808) & (37{,}729)& (46{,}966) \\
\multirow{2}{*}{$\pi_2$}& 3.74377  & 3.94392 & 3.94214 & 5.03094 & 3.94568 & 3.81509 & 3.92345\\
& (50{,}000) & (44{,}850) & (44{,}893) & (27{,}144) & (44{,}814) & (48{,}075) & (45{,}358) \\
\multirow{2}{*}{$\pi_3$}& 6.77106 & 5.90052  & 5.91950 & 4.90988 & 5.90241 & 5.99899 & 6.31881\\
& (50{,}000) & (66{,}416) & (65{,}978) & (96{,}946) & (66{,}378) & (64{,}196) & (57{,}676) \\
\hline
$\mathcal{L_{\mathrm{ave}}}$& 4.40273 & \textbf{4.30450} & 4.30488 & 4.57807 & 4.30482 & 4.30868 & 4.34091\\
\end{tabular}
}
\end{table}

For this transdimensional sampling example, it also makes sense to
consider the fixed dimension function of interest methods of
\cite{sisson_fan07}, using the mean intensity function of the Poisson
process or the distance to nearest changepoint, each evaluated at 100
equally spaced grid points on $[0,1]$. The Monte Carlo variances used
in these strategies estimate the variances displayed in the plots of
Fig. \ref{fig:fois} at the reference points, divided by the current
sample size. The performance of these fixed dimensional strategies is
particularly poor under the loss function
$\mathcal{L_{\mathrm{max}}}$. Importantly, it should also be noted
that the sample sizes and performance vary considerably depending upon
which of the two arbitrary functions of the reference points are used.

\subsection{IEEE VAST 2008 Challenge Data}\label{sec:vast_data}
This final example now illustrates how the method performs in the
presence of a much larger number of target distributions, in the
context of network security. The IEEE VAST 2008 Challenge data are
synthetic but realistically generated records of mobile phone calls
for a small community of 400 individuals over a ten day period. The
data can be obtained from
{\url{www.cs.umd.edu/hcil/VASTchallenge08}}. The aim of the original
challenge was to find anomalous behaviour within this social network,
which might be indicative of malicious coordinated activity.

One approach to this problem is to monitor the call patterns of
individual users and detect any changes from their normal behaviour,
with the idea that a smaller subset of anomalous individuals will then
be investigated for community structure. In particular, this approach
has been shown to be effective with these data when monitoring the
event process of incoming call times for each individual
\citep{heard10}. After correcting for diurnal effects on normal
behaviour, this approach can be reduced to a changepoint analysis of
the intensities of 400 Poisson processes of the same character as
Section \ref{sec:results_multivariate}. For the focus of this article,
it is of interest to see how such an approach could be made more
feasible in real time by allocating computational resource between
these 400 processes more efficiently.

\begin{figure}[ht!]
\centering
\makebox{\includegraphics[height=6.5cm]{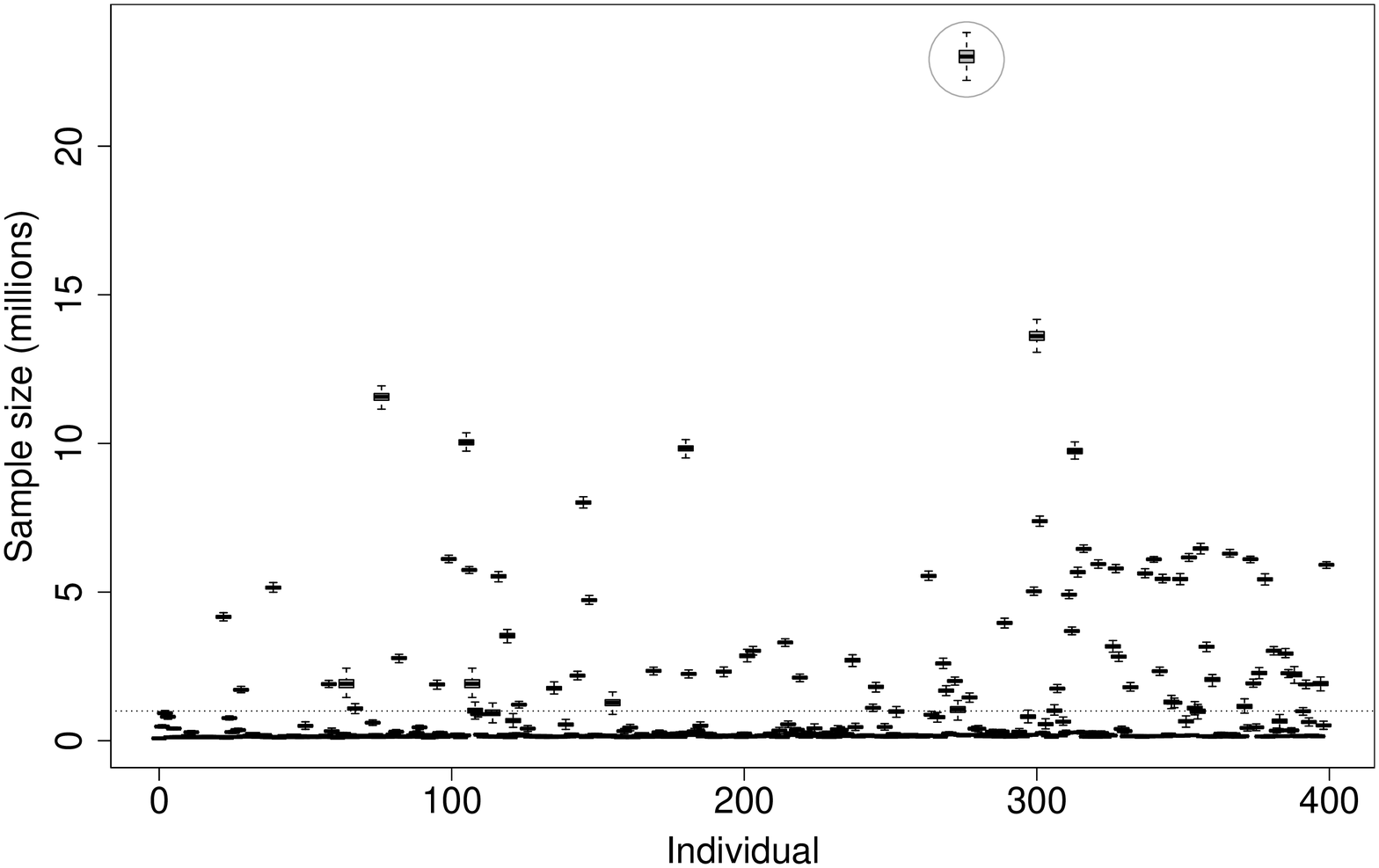}}
\makebox{\includegraphics[height=6.5cm]{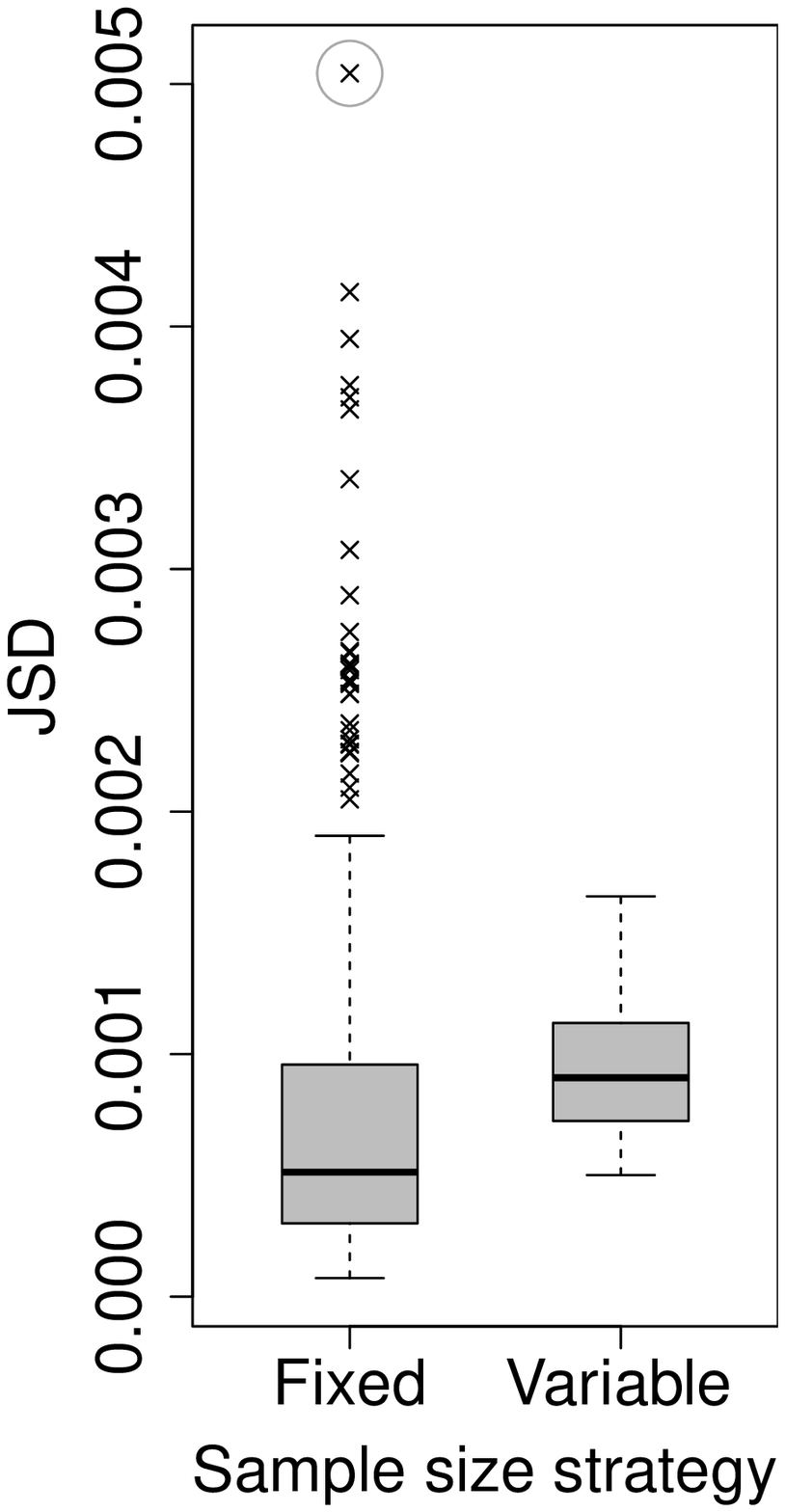}}\\
\caption{\label{fig:vast}Results of VAST data analysis. Left:
  Distribution of sample sizes under the
  Grassberger strategy for each individual posterior. Right:
  Distribution across individuals of estimated Monte Carlo
  divergence error under fixed or variable (Grassberger) sample size
  strategies.}
\end{figure}

Fig. \ref{fig:vast} shows the contrasting performance between an equal
computational allocation of one million Markov chain Monte Carlo
samples to each process against the variable sample size approach
using Grassberger's entropy bias estimate, for the same total
computational resource of 400 millions samples and using the loss
function $\mathcal{L_{\mathrm{max}}}$. The left hand plot shows the
distribution of sample sizes for the individual processes over $M=200$
repetitions, using 5{,}000 initial samples and an average allocation
of one million samples for each posterior target; the dashed line
represents the fixed sample size strategy. The sample sizes vary
enormously across individuals. However, for each individual the
variability between runs is much lower, showing that the method is
robust in performance. The right hand plot shows the resulting Monte
Carlo divergence errors of the estimated distributions from the
targets. Ideal performance under $\mathcal{L_{\mathrm{max}}}$ would
have each of these errors approximately equal, and the variable sample
size method gets much closer to this ideal. The circled case in the
right hand plot indicates the process which has the highest error when
using a fixed sample size, and this corresponds to the same individual
process that always gets the highest sample size allocation under the
adaptive sample size strategy in the left hand plot. This individual
has a very changeable calling pattern, suggesting several possible
changepoints: no calls in the first five days, then two calls one hour
apart, then another two days break, and then four calls each day for
the remainder of the period.

\section{Discussion}\label{sec:discussion}

It was remarked in the review paper of \cite{sisson2005} on
transdimensional samplers that ``a more rigorous default assessment of
sampler convergence'' than the existing technology is required, and
this has remained an open problem. This article is a first step
towards establishing such a default method from a decision theoretic
perspective, proposing a framework and methodology which are
rigorously motivated and fully general in their applicability to all
distributional settings.

Note that when the samplers induce autocorrelation, which is
commonplace with Metropolis-Hastings (MH) Markov chain Monte Carlo
simulation, then the decision rule \eqref{eq:ave_rule} for
$\mathcal{L_{\mathrm{ave}}}$ becomes more complicated since
independence was assumed in the derivation of \eqref{eq:dehat_KL}. If
one or more of the samplers has very high serial autocorrelation, then
drawing additional samples from those targets will become less
attractive under $\mathcal{L_{\mathrm{ave}}}$, as with high
probability very little extra information will be obtained from the
next draw. It is still possible to proceed in this setting by adapting
\eqref{eq:dehat_KL} to admit autocorrelation; for example, the
rejection rate of the Markov chain could be used to approximate the
probability of observing the same bin as the last sample, and
otherwise draws could be assumed to be more realistically drawn from
the target. However, for reasons of brevity this is not pursued
further in this work, and of course the efficacy would depend entirely
on the specifics of the MH/other sampler. Importantly, this issue
should not be seen as a decisive limitation of the proposed
methodology when using $\mathcal{L_{\mathrm{ave}}}$, since although
thinning was used in the Markov chain Monte Carlo examples of Sections
\ref{sec:results_multivariate} and \ref{sec:vast_data} to obtain the
next sample for use in calculating the convergence criteria, this
would not prevent the full sample from being retained and utilised
without thinning for the actual inference problem. The amount of
thinning could be varied between samplers if appropriate, and this
could be counterbalanced by weighting the errors in
\eqref{eq:ave_loss} accordingly.

Another related problem which could be considered is that of
importance sampling. If samples cannot be obtained directly from the
target $\pi$ but instead from some importance distribution with the
same support, then it would be useful to understand how these error
estimates and sample size strategies can be extended to the case where
the empirical distribution of the samples has associated weights. In
addressing the revised question of how large an importance sample
should be, there should be an interesting trade-off between the
inherent complexity of the target distributions, which has been the
subject of this article, and how well the importance distributions
match those targets.

\bibliographystyle{Chicago}

\end{document}